\documentclass[a4paper,11pt]{article}
\pdfoutput=1 	% if your are submitting a pdflatex (i.e. if you have
\usepackage{jheppub} 	% for details on the use of the package, please
\usepackage{verbatim}
\usepackage{graphicx}	% Include figure files
\usepackage{dcolumn}	% Align table columns on decimal point
\usepackage{bm}		    % bold math
\usepackage{hyperref}
\usepackage{amsmath}
\usepackage{amssymb}
\usepackage{verbatim}
\usepackage{slashed}
\usepackage{hyperref}
\usepackage{mathtools}
\usepackage{tabularx}
\usepackage{pifont}
\usepackage{subcaption}
\usepackage{tikz}
\usepackage{tikz-feynman}

\newcommand{\cmark}{\ding{51}}%
\newcommand{\xmark}{\ding{55}}%

\preprint{IBS-CTPU-26-19}

\title{\boldmath  Probing Neutrinophilic Axion-Like Particles in Tritium Beta Decay}

\author[a]{Yi Chung,}
\author[b]{Florian Goertz,}
\author[b]{Maya Hager,}
\author[c]{and Joscha Lauer}

\affiliation[a]{
Particle Theory and Cosmology Group, Center for Theoretical Physics of the Universe, \\Institute for Basic Science (IBS), Daejeon, 34126, Korea
}
\affiliation[b]{
Max-Planck-Institut für Kernphysik, Saupfercheckweg 1, 69117 Heidelberg, Germany
}
\affiliation[c]{
Institute for Astroparticle Physics (IAP), Karlsruhe Institute of Technology (KIT), Hermann-von-Helmholtz-Platz 1, 76344 Eggenstein-Leopoldshafen, Germany}

\emailAdd{yichung886@ibs.re.kr}
\emailAdd{florian.goertz@mpi-hd.mpg.de}
\emailAdd{maya.hager@mpi-hd.mpg.de}
\emailAdd{joscha.lauer@kit.edu}

\abstract{We investigate the prospects for constraining neutrinophilic axion-like particles via measurements of the tritium beta decay spectrum, as performed by the KATRIN experiment and its planned TRISTAN detector upgrade. We study in detail the resulting spectral modifications and the corresponding experimental sensitivity. The relevant complementary searches are also discussed for comparison and we derive the most up-to-date and robust constraints on keV-scale neutrinophilic axion-like particles, covering both lepton-number-conserving and lepton-number-violating interactions. Cosmological constraints are generally more stringent; however, this conclusion relies on the assumption that the particles remain unchanged from the early universe to the present day. We therefore construct a model in which the neutrinophilic axion-like particle, being a pseudo-Nambu-Goldstone boson of an extended scalar sector, emerges from a spontaneous symmetry breaking featuring a non-trivial thermal history. We show that the model naturally evades the conventional cosmological bounds, allowing tritium beta decay measurements to provide the leading constraints.}

\begin{document}
\maketitle
\flushbottom

\section{Motivation and setup}

Axion-like particles (ALPs) are light pseudoscalar pseudo-Nambu-Goldstone bosons (pNGBs) of a spontaneously broken approximate global symmetry with their interactions controlled by the resulting approximate shift symmetry. They arise in many extensions of the Standard Model (SM) of Particle Physics and can play an important role in addressing unresolved questions of nature. They can provide for example viable dark matter candidates~\cite{Preskill:1982cy,Abbott:1982af,Dine:1982ah} or can serve as mediators between visible and dark sectors~\cite{Beacham:2019nyx,Dror:2023fyd,Fitzpatrick:2023xks,Armando:2023zwz}. Moreover, if the spontaneously broken symmetry features an appropriate color anomaly, they can solve the strong CP problem by dynamically relaxing the total CP-violating angle to zero. Interestingly, they also emerge in solutions to the flavor puzzle~\cite{Ema:2016ops,Calibbi:2016hwq}, the hierarchy problem ~\cite{Graham:2015cka,Choi:2015fiu,Gripaios:2009pe,Gherghetta:2020ofz,Goertz:2026oco}, or the puzzle of small neutrino masses~\cite{Chikashige:1980ui,Gelmini:1980re,Schechter:1981cv}.
Finally, ALPs are a generic prediction of string compactifications~\cite{Svrcek:2006yi,Arvanitaki:2009fg,Cicoli:2012sz}. In general, they offer a fascinating window to high scales via light new states, with their mass being protected by the shift symmetry.
Besides cosmological and astrophysical probes, their light masses make them accessible to a variety of precision experiments, making them promising targets for these laboratory searches.

While most searches for ALPs focus on their couplings to photons, gluons, and charged fermions, a less explored direction is the search for ALPs via their interactions with neutrinos, which is experimentally challenging. Here, experiments analyzing the electron spectrum in tritium beta decays, where $\mathrm{^3H} \rightarrow \mathrm{^3He^+} + e^- + \bar{\nu}_e$, such as KATRIN and its planned TRISTAN detector upgrade \cite{Aker_2021, Aker_2022, doi:10.1126/science.adq9592, PhysRevLett.134.251801, KATRIN_Collaboration2025-zx, Mertens_2021, 2022ConceptualDR, acharya2026katrinsensitivitykevsterile, Lauer:2024whe}, offer an interesting possibility to find neutrinophilic ALPs ($a$). If their mass resides below the spectral endpoint $E_0\approx 18.6$\,keV, a characteristic change in the beta spectrum emerges, as now $\mathrm{^3H} \rightarrow \mathrm{^3He^+} + e^- + \bar{\nu}_e + a$ becomes possible, which allows to explore uncharted territory. In this paper, we will provide projected limits of a TRISTAN-like setup on the ALP--neutrino coupling in dependence on the ALP mass. In particular, we will take into account complementary searches and cosmological constraints and show that the most severe bounds can be avoided in our ALP model, such that beta decay experiments can set competitive limits.

The relevant Lagrangian, describing a neutrinophilic ALP, reads
\begin{align}\label{ALP}
\mathcal{L}_{\rm ALP}=\frac{1}{2}(\partial_\mu a)(\partial^\mu a)-\frac{1}{2}m_a^2 a^2
+\frac{\partial_\mu a}{f_a}\,\bar{\nu}\gamma^\mu\gamma_5 \nu
~,
\end{align}
where the interaction respects the shift symmetry of the ALP, and $f_a$ is the new physics scale associated with the symmetry breaking, which determines the strength of the interaction. To match with common phenomenological analyses, we rewrite the interaction term as
\begin{align}\label{int}
\mathcal{L}_{\rm int.}=ig_\nu a\bar{\nu}\gamma_5\nu \quad
{\rm with}\quad
g_\nu=\frac{M_\nu^a}{f_a}~,
\end{align}
where $M_\nu^a$ denotes the component of the neutrino mass associated with the new physics scale $f_a$ (not necessary the exact neutrino mass), determining the size of the ALP--neutrino coupling $g_\nu$.
Here, we consider the neutrinophilic ALP to couple to the lepton number conserving current, featuring light right-handed (RH) neutrinos that carry the same lepton number as the left-handed (LH) neutrinos. However, the ALP can also couple to a lepton number violating current $\overline{\nu^c}\gamma_5 \nu$ with coupling $g_{\nu,\slashed{L}}$. %In the scenario of Majorana neutrinos, where the neutrino is its own anti-particle ($\nu=\nu^c$), the distinction disappears, and both bilinears reduce to the same low-energy operator.
A well known example is the Majoron \cite{Chikashige:1980ui,Gelmini:1980re,Schechter:1981cv}, which is the pNGB of broken lepton number, with its scale and couplings typically related to the seesaw scale \cite{Minkowski:1977sc,Gell-Mann:1979vob,Yanagida:1979as,Glashow:1979nm,Mohapatra:1979ia}. 
For tritium beta decay and most complementary searches, there is no difference between the two cases so we will just give the constraints in terms of $g_\nu$. Particular observables such as neutrinoless double beta decay can put additional bounds, which only apply to the $g_{\nu,\slashed{L}}$ coupling. As we will discuss later in more detail, due to the strict constraints from neutrinoless double beta decay measurements, the lepton number conserving case will be more relevant for our study.

As mentioned, such light new scalars as envisioned here are subject to strong complementary constraints, in particular from cosmology. Indeed, given they remain relativistic during Big Bang Nucleosynthesis (BBN), they lead to too much additional radiation, i.e. a too large $N_{\rm eff}$, which is experimentally excluded unless the coupling to neutrinos is tiny, orders of magnitude below the prospective sensitivity of KATRIN, which seems to render corresponding searches for light scalars hopeless.
However, this conclusion only holds if the characteristics of the scalar (and/or neutrinos) remain unchanged from the early to the late universe. In particular for ALPs, which emerge from spontaneous symmetry breaking, this is not a given. In the early universe the symmetry is expected to be restored and the scalar will have a mass at the order of the scale $f_a$, while after a symmetry-breaking phase transition, the light ALP emerges. If this transition happens after BBN, the strong bounds from $N_{\rm eff}$ can be avoided. Thus, ALPs are arguably among the most motivated targets for light-scalar searches in tritium beta decay. In this work, we will construct a corresponding explicit scenario avoiding cosmological constraints, such that indeed these experiments could lead to meaningful leading constraints on such light new scalars.

The paper is organized as follows. In Sec.~\ref{sec:KATRIN} we compute the effect of neutrinophilic ALP emission on the tritium beta spectrum and provide the resulting statistical sensitivity on $g_\nu$ at a TRISTAN-like setup. In Sec.~\ref{sec:Other} we discuss the most relevant complementary searches in lab experiments and in astrophysics and cosmology and give the resulting bounds. Subsequently, in Sec.~\ref{sec:Model} we present a UV completion for the neutrinophilic ALP that evades the most constraining cosmological bounds via a non-trivial thermal history, to de discussed in detail, while in Sec.~\ref{sec:other} we analyze potential additional signatures due to the UV setup. Finally, in Sec.~\ref{sec:Conclusion} we conclude.

\section{Sensitivity of tritium beta decay}\label{sec:KATRIN}

High-precision measurements of the tritium beta decay spectrum, enabled by the KATRIN experiment and its upcoming TRISTAN detector upgrade, provide remarkable sensitivity to neutrino-related new physics scenarios \cite{KATRIN_Collaboration2025-zx, PhysRevLett.134.251801}. In this section, we compute the spectral effect of neutrinophilic ALP emission in tritium beta decay 
and derive the resulting statistical sensitivity to the coupling $g_\nu$ for a TRISTAN benchmark configuration, thereby establishing the reach of such measurements for this class of new physics.

\subsection{Spectral modifications from neutrinophilic ALP emission}\label{sec:spectrum}

Modifications of the electron energy spectrum in tritium beta decay due to the additional emission of a real light neutrinophilic pseudoscalar were first presented in \cite{Arcadi:2018xdd}. In the computation of the spectrum, the authors apply an exact treatment of the relativistic four-body kinematics in the decay $\mathcal{A} \rightarrow \mathcal{B} + e^- + \bar{\nu}_e + a$, which is shown in Fig.~\ref{f:real_emission_nu}. For applicability in their analysis, the authors reduce the numerical result from the exact treatment by fitting it to a fixed power law in a kinematic variable. We compute the spectrum within the recoil approximation, which yields a simple closed-form analytical result, avoiding both the numerical integration required by exact four-body kinematics in \cite{Arcadi:2018xdd} and the loss of functional information incurred by the subsequent fit:\begin{figure}[tb]
    \centering
    %------------ Subfigure B --------------%
    \begin{subfigure}{0.32\textwidth}
        \centering
        \begin{tikzpicture}
            
            \begin{feynman}
                \vertex (n)  at (-1.50,  0.00) {$\mathcal{A}$};
                \vertex (v)  at ( 0.00,  0.00);
                \vertex (p)  at ( 0.75, -1.00) {$\mathcal{B}$};
                \vertex (e)  at ( 2.0,  0.75) {\(e^-\)};
                \vertex (nu) at ( 2.00, -0.50) {\(\bar{\nu}_e\)};
                \diagram*{
                    (n)  -- [plain] (v),
                    (v)  -- [plain] (p),
                    (v)  -- [postaction={decorate},
                                decoration={markings, mark=at position 0.65 with {\arrow{Latex[length=6pt,width=4pt]}}}] (e),
                    (nu)  -- [postaction={decorate},
                                decoration={markings, mark=at position 0.5 with {\arrow{Latex[length=6pt,width=4pt]}}}] (v),
                };
            \end{feynman}
            
            \fill[white] (0,0) circle (0.2);
            \begin{scope}
                \clip (0,0) circle (0.2);
                \foreach \i in {-1.0, -0.9, ..., 1.0} {
                    \draw[thin] (\i - 0.5, -0.5) -- (\i + 0.5, 0.5);
                }
            \end{scope}
            \draw[thick] (0,0) circle (0.2);
            
        \end{tikzpicture}
        \caption{SM beta decay.}\label{f:beta_decay}
    \end{subfigure}
    %
    %------------ Subfigure B --------------%
    \begin{subfigure}{0.32\textwidth}
        \centering
        \begin{tikzpicture}
            
            \begin{feynman}
                \vertex (n)  at (-1.50,  0.00) {$\mathcal{A}$};
                \vertex (v)  at ( 0.00,  0.00);
                \vertex (p)  at ( 0.75, -1.00) {$\mathcal{B}$};
                \vertex (e)  at ( 2.00,  0.75) {$e^-$};
                \vertex (vb) at ( 1.00, -0.250);
                \vertex (nu) at ( 2.00, -0.875) {$\bar{\nu}_e$};
                \vertex (b)  at ( 2.00,  -0.125) {$a$};
                \diagram*{
                    (n)  -- [plain]   (v),
                    (v)  -- [plain]   (p),
                    (v)  -- [postaction={decorate},
                                decoration={markings, mark=at position 0.65 with {\arrow{Latex[length=6pt,width=4pt]}}}] (e),
                    (nu)  -- [postaction={decorate},
                                decoration={markings, mark=at position 0.65 with {\arrow{Latex[length=6pt,width=4pt]}}}] (vb),
                    (vb)  -- [postaction={decorate},
                                decoration={markings, mark=at position 0.5 with {\arrow{Latex[length=6pt,width=4pt]}}}] (v),
                    (vb) -- [scalar]   (b),
                };
            \end{feynman}
            
            \fill[white] (0,0) circle (0.2);
            \begin{scope}
                \clip (0,0) circle (0.2);
                \foreach \i in {-1.0, -0.9, ..., 1.0} {
                    \draw[thin] (\i - 0.5, -0.5) -- (\i + 0.5, 0.5);
                }
            \end{scope}
            \draw[thick] (0,0) circle (0.2);
            
        \end{tikzpicture}
        \caption{ALP emission off $\bar{\nu}_e$.}\label{f:real_emission_nu}
    \end{subfigure}

    \caption[Beta decay with neutrinophilic ALP]{Beta decay of $\mathcal{A}$ with a neutrinophilic ALP $a$. (\subref{f:beta_decay}) shows the SM process; (\subref{f:real_emission_nu}) shows the four-body channel, where the ALP is emitted off the neutrino as a real particle.}\label{f:decays}
\end{figure}
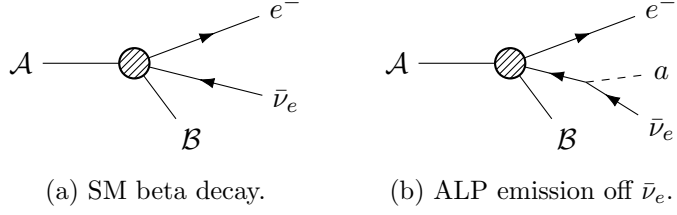 Based on the large absolute mass $m_{\mathcal{A}}$ of molecular tritium, in contrast to a comparably small mass difference $m_{\mathcal{A}}-m_{\mathcal{B}}=\Delta \approx E_0+m_e$, which reflects the released decay energy, we neglect small terms suppressed by $\Delta/m_{\mathcal{A}} < 10^{-4}$. The SM beta decay process is illustrated in Fig.~\ref{f:beta_decay}. In the recoil approximation, the corresponding differential beta spectrum as a function of the electron kinetic energy $E$ assumes the well-known form
\begin{align}
    \frac{\mathrm{d} \Gamma_\beta}{\mathrm{d} E} =&\, \frac{G_{\mathrm{F}}^2 \lvert V_{ud}\rvert^2 (g_V^2+ 3 g_A^2)}{2 \pi^3}\label{e:beta_spectrum}
    \\
    &\quad \cdot (E+m_e) \sqrt{(E+m_e)^2-m_e^2}\,\underbrace{(E_0-E)}_{\epsilon} \underbrace{\sqrt{(E_0-E)^2-\smash{m_{\nu}^2}}}_{P_\epsilon(m_{\nu}^2)} \cdot\, F(E,Z+1).
    \nonumber
\end{align}
Here, $g_V$ and $g_A$ are the vector and axial-vector coupling constants of the nuclear transition. In the SM process, the remaining energy $\epsilon$ is carried away by the neutrino, such that $\epsilon$ and $P_\epsilon(m_{\nu}^2)$ define the energy $E_\nu$ and momentum $P_\nu$ of the neutrino. The Fermi function\footnote{At low $Z$, as in tritium decay with a daughter charge of $Z=2$, the Fermi function is approximated by $F(E, Z) = \frac{2\pi y}{1-\exp(-2\pi y)}$. Here, $y=Z\alpha \,(E+m_e)/|\vec{p}_e|$, with the fine structure constant $\alpha$.} $F(E, Z)$ accounts for the Coulomb interaction between the outgoing electron and the daughter nucleus.

Including additional emission of the (on-shell) ALP off the neutrino, governed by the interaction Lagrangian \eqref{int} and the tree-level structure from Fig.~\ref{f:real_emission_nu}, we find the real emission channel spectrum
\begin{align}
    \frac{\mathrm{d} \Gamma_a^{\mathrm{real}}}{\mathrm{d} E}(m_{\nu}=0) &= \frac{g_\nu^2}{(4 \pi)^2}\frac{1}{8} \left(\left(15 - 6 \beta_a^2 - \beta_a^4\right)\operatorname{artanh}(\beta_a) - \left( 15 - \beta_a^2 \right) \beta_a \right) \frac{\mathrm{d} \Gamma_\beta}{\mathrm{d} E}(m_{\nu}=0),\nonumber\\
    &\quad\,\, \text{with}\quad \beta_a  = \frac{P_\epsilon(m_a^2)}{\epsilon}= \sqrt{1 - \left( \frac{m_a}{\epsilon} \right)^2}.\label{e:real_emission}
\end{align}
In this case, $\epsilon$ from Eq.~\eqref{e:beta_spectrum} describes the total energy available in the system of ALP and neutrino, and $\beta_a$ tracks the maximum velocity of the ALP. Our result represents the four-body spectrum within the recoil approximation, and vanishes beyond ${E = E_0 -m_a}$. The small neutrino mass, constrained to $m_{\nu}<0.45\,\mathrm{eV}$ (90\% C.L.) by the KATRIN experiment~\cite{doi:10.1126/science.adq9592}, can be effectively neglected ($m_{\nu}=0$) for our purposes.

While the real emission develops a logarithmic divergence due to collinear emission for $m_a \rightarrow 0$ (for $m_{\nu}=0$), after we include the virtual correction from the one-loop neutrino self-energy, due to an ALP loop, the inclusive result is IR-finite, see also \cite{Dev:2024ygx}. This virtual correction is accounted for in the renormalization factor $\sqrt{\mathcal{Z}}$, which combines the neutrino field-renormalization constant and the LSZ residue of the renormalized propagator. The UV-divergent part of $\mathcal{Z}$ is associated with neutrino wave-function renormalization and is absorbed by the renormalization of the Wilson coefficient $C_\beta(\mu)$ of the effective weak four-fermion operator, entering \eqref{e:beta_spectrum} as a non-trivial (squared) prefactor at one-loop order. In the $\overline{\text{MS}}$ scheme, as in \cite{Dev:2024ygx}, the finite part $\mathcal{Z}_{\mathrm{fin}}(\mu)$ of $\mathcal{Z}$ modifies the normalization of the SM channel according to
\begin{align*}
\mathrm{d}\Gamma_\beta \rightarrow \mathcal{Z}_{\mathrm{fin}}(\mu)\,\mathrm{d}\Gamma_\beta(\mu), \quad \mathcal{Z}_{\mathrm{fin}}(m_\nu=0, \mu) &= 1 -\frac{g_\nu^2}{(4\pi)^2}\left(\ln\!\left(\frac{\mu}{m_a}\right)+\frac{1}{4}\right),
\end{align*}
where $\mu$ dependence enters $\mathrm{d}\Gamma_\beta(\mu)$ through $C_\beta(\mu)$. Summing the renormalized SM-like channel~\eqref{e:beta_spectrum} and the real emission branch \eqref{e:real_emission} yields the total spectrum for $m_\nu = 0$, which is independent of $\mu$ (at $\mathcal{O}(g_\nu^2)$). The ALP remains unresolved
in this inclusive observable. Without affecting the result at the considered order, i.e., up to $g_\nu^2$, we therefore fix the renormalization scale $\mu$ to the characteristic scale $E_0$. The full result is
\begin{align}\label{e:total_spectrum}
\frac{\mathrm{d}\Gamma}{\mathrm{d}E} &= \left[1+\frac{g_\nu^2}{(4\pi)^2}\frac{1}{8}\left(\left(15-6\beta_a^2-\beta_a^4\right)\operatorname{artanh}(\beta_a)-\left(15-\beta_a^2\right)\beta_a-8\ln\!\left(\frac{E_0}{m_a}\right)-2\right)\right]\frac{\mathrm{d}\Gamma_\beta}{\mathrm{d}E},
\end{align}
where $C_\beta(\mu = E_0)$ is implicitly included in $\mathrm{d}\Gamma_\beta$ as an overall normalization effect; since our analysis relies exclusively on the spectral shape, 
$C_\beta(\mu)$ cancels and does not need to be determined independently. The logarithmic contribution in $\mathcal{Z}_{\mathrm{fin}}(\mu)$ exactly cancels the collinear divergence of the real emission, ensuring an IR-finite prediction for $m_a \rightarrow 0$, consistent with the KLN theorem \cite{KINOSHITAMassSingularitiesFeynman1962, LEEDegenerateSystemsMass1964}.\begin{figure}
    \centering
    \includegraphics[width=0.8\textwidth]{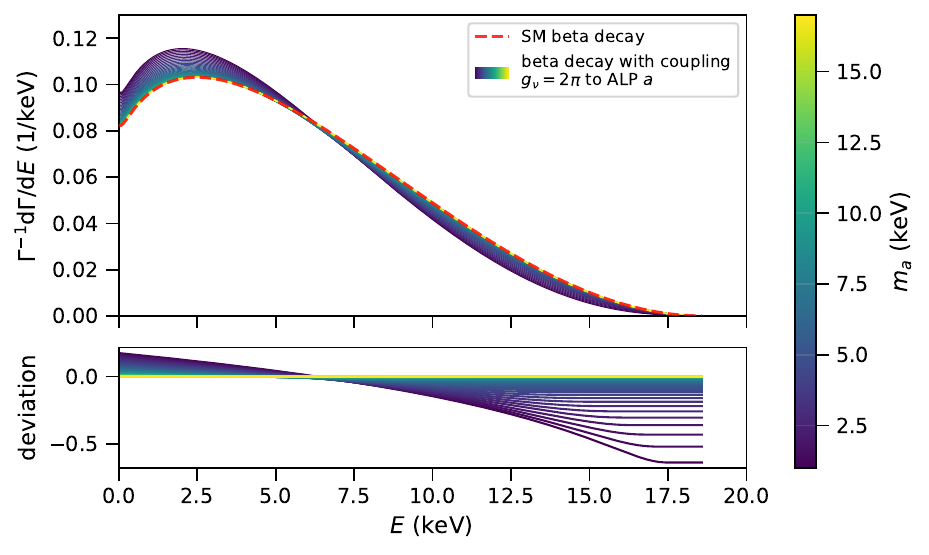}
    \caption[Tritium beta spectrum with emission of neutrinophilic ALPs]{Tritium beta spectrum with emission of a neutrinophilic ALP $a$. The normalized beta spectrum $\Gamma^{-1}\,\mathrm{d} \Gamma/\mathrm{d} E$ (Eq.~\eqref{e:total_spectrum}) as a function of the electron kinetic energy $E$ is shown for different keV ALP masses $m_a$, indicated by the color (upper panel). A large coupling $g_\nu = 2 \pi$ is assumed to improve visibility. The dashed red line corresponds to the SM beta decay (Eq.~\eqref{e:beta_spectrum}). In the lower panel, the relative difference between the ALP spectrum and the SM prediction is plotted.}
    \label{fig:ALP_spectrum}
\end{figure}

In Fig.~\ref{fig:ALP_spectrum}, the normalized modified beta spectrum $\Gamma^{-1}\,\mathrm{d} \Gamma/\mathrm{d} E$ of \eqref{e:total_spectrum} is plotted as a function of the electron kinetic energy $E$ for different keV-range ALP masses $m_a$ and a large fixed coupling of $g_\nu = 2 \pi$. For $m_a \rightarrow E_0$ the ALP contribution ceases, such that the SM spectrum \eqref{e:beta_spectrum} is recovered. In the lower panel, the relative deviation is shown, revealing the underlying structure of the spectral distortion due to ALP emission. In the following, we investigate how this shape effect translates to sensitivity to neutrinophilic ALPs.

\subsection{Statistical sensitivity}\label{sec:sensitivity}

The KATRIN experiment probes the endpoint region of the tritium beta spectrum, up to several tens of eV below $E_0$, where the sensitivity to the neutrino mass in \eqref{e:beta_spectrum} is maximal. The windowless gaseous molecular tritium source of the experiment provides a stable, high activity of up to $10^{11}\,\mathrm{Bq}$. Employing magnetic adiabatic collimation with an electrostatic (MAC-E) filter, KATRIN measures the highpass-filtered integral beta spectrum, corresponding to the electron count rate integrated above an adjustable retarding energy threshold. A spectral model, convolved with the experimental response function, is fitted to the measured spectrum to search for spectral distortions originating from the neutrino mass or BSM physics \cite{Aker_2021, Aker_2022, doi:10.1126/science.adq9592, KATRIN_Collaboration2025-zx, PhysRevLett.134.251801}. The sensitivity of KATRIN to neutrinophilic pseudoscalars and other types of new light bosons is addressed in \cite{Arcadi:2018xdd, Aker_2022, Lauer:2024whe}.

After reaching its goal of 1000 measurement days in 2025, KATRIN transitions to the TRISTAN phase. In this phase, the experimental reach is extended to keV sterile neutrinos through integration of the TRISTAN detector \cite{2022ConceptualDR, Mertens_2021, acharya2026katrinsensitivitykevsterile}. In contrast to the integral endpoint measurement, the differential beta spectrum is directly measured, covering the full electron energy range accessible in the decay. The expected measured rate of electrons is then ${\sim\,}5 \cdot 10^7\,\mathrm{cps}$, compared to an average count rate of ${\sim\,}2.4\,\mathrm{cps}$ in the integral endpoint measurement without the new detector. The TRISTAN detector modules are optimized for this differential operation mode at rates as high as ${\sim\,}100\,\mathrm{kcps}$ per pixel, and reach an energy resolution of better than $300\,\mathrm{eV}$ FWHM in response to $20\,\mathrm{keV}$ electrons.

In the previous analysis \cite{Arcadi:2018xdd}, the statistical sensitivity was derived using a semi-analytical parametrization of the spectrum with exact kinematics. Our analysis improves on this in two key aspects: (1) although the kinematic treatment in the preceding work is exact, it is subsequently reduced to a fixed power law in $\epsilon - m_a$ for the sensitivity study, discarding functional information; our closed-form result \eqref{e:real_emission}, obtained within the recoil approximation to the same order as the beta spectrum \eqref{e:beta_spectrum}, retains the full energy dependence of the spectrum without this loss; and (2) we base our sensitivity study on more realistic predictions for the experimental event statistics and the electron energy interval considered in the analysis, superseding the more optimistic assumptions underlying the earlier work.

We derive a purely statistical sensitivity based on the spectrum \eqref{e:total_spectrum}, obtained in the recoil approximation to the same order as the SM beta spectrum \eqref{e:beta_spectrum}. To estimate the statistical reach of the TRISTAN setup regarding neutrinophilic ALPs, we adopt the estimated amount of $N_{\mathrm{tot}} = 4 \times 10^{14}$ electrons in the range of electron kinetic energies between $1.5\,\mathrm{keV}$ and $E_0$, corresponding to a
data-taking period of four months. The total number of counts is fixed to $N_{\mathrm{tot}}$, such that only shape information enters the analysis. We also keep both the endpoint $E_0 = 18.6\,\mathrm{keV}$ and the neutrino mass $m_\nu=0$ fixed, as their effect is negligible for the full spectrum fitting. To isolate the purely statistical reach of tritium beta decay from experiment-specific effects, we do not include background and energy resolution. However, we confirmed that the contribution of solely these two effects in a TRISTAN-like setup \cite{acharya2026katrinsensitivitykevsterile} has negligible impact on the sensitivity.

The remaining free parameters $\theta=(m_a, g_\nu)$ of our model are the ALP mass $m_a$ and its coupling $g_\nu$, and $\theta_0$ defines the parameter set of the null hypothesis, coinciding with the SM beta spectrum \eqref{e:beta_spectrum} in our analysis range (either by $g_\nu = 0$ or by $m_a>E_0$). We employ the $\chi^2$ test statistic \cite{Cowan2011-pt} derived from the binned likelihood ratio in the high-count Gaussian approximation
\begin{align}
    \chi^2(\theta) =&\, \sum_{i} \frac{(N_i(\theta)-N_i(\theta_0))^2}{N_i(\theta_0)},\label{e:chi_square}\\ &\,N_i(\theta)=\frac{N_{\mathrm{tot}}}{\sum_j n_j(\theta)}\,n_i(\theta),\quad n_i(\theta)=\int_{E_i^{\mathrm{min}}}^{E_i^{\mathrm{max}}}\frac{\mathrm{d} \Gamma(\theta)}{\mathrm{d} E}\,\mathrm{d}E,\nonumber
\end{align}
where $\mathrm{d}\Gamma(\theta)$ is defined by \eqref{e:total_spectrum}. The sum runs over 50 equidistant electron energy bins.\begin{figure}
    \centering
    \includegraphics[width=0.8\textwidth]{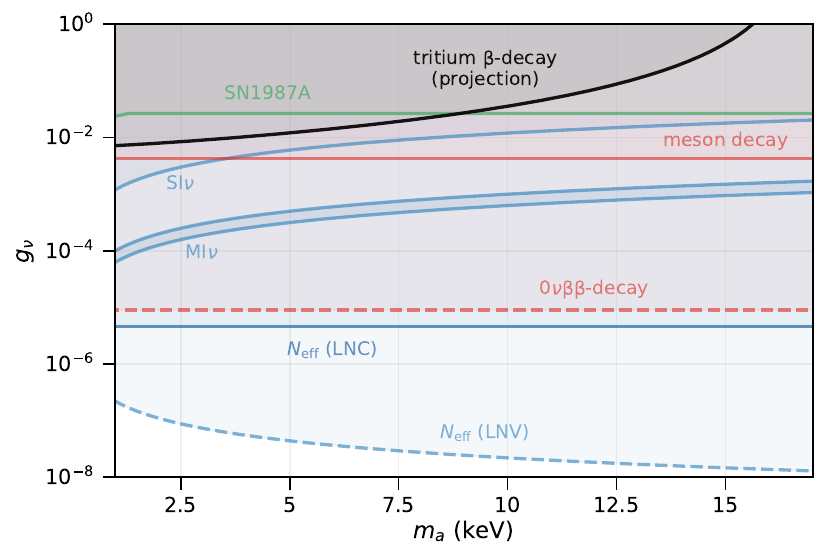}
    \caption[Tritium beta decay sensitivity]{Sensitivity of tritium beta decay and other probes to neutrinophilic ALPs. The boundary of the accessible coupling $g_\nu$ is shown as a function of the ALP mass $m_a$. For tritium beta decay, the resulting purely statistical sensitivity (90\%\,C.L.) of the TRISTAN benchmark setup (this work) is displayed in black. Other bounds from complementary searches (Sec.~\ref{sec:Other}) are shown in blue, green, and red for cosmological, astrophysical, and laboratory bounds, respectively. Dashed lines indicate the constraints from processes with lepton number violation (LNV).} 
    \label{fig:sensitivity_with_other_bounds}
\end{figure}  Our test statistic is evaluated on an unfluctuated Asimov dataset $\{ N_i(\theta_0) \}$ generated under the null hypothesis.

The analysis is performed on a discrete two-dimensional grid of $\theta$, similar to \cite{acharya2026katrinsensitivitykevsterile, PhysRevLett.134.251801, Lauer:2024whe}. It spans $10^3 \times 10^3$ points, equidistant in ${m_a \in[0.001, 16.5]\,\mathrm{keV}}$, and logarithmically spaced in ${g_\nu \in [0.004, 10]}$. With $\chi^2(\theta_0)=0$, parameters $\theta$ where $\chi^2(\theta)=\chi_{\mathrm{crit}}^2$ represent the statistical sensitivity at a confidence level of $\mathrm{C.L.}=\mathcal{P}(\chi_k^2 \leq \chi_{\mathrm{crit}}^2)$, according to Wilks' theorem. $\chi_k^2$ is the chi-squared distribution with $k$ degrees of freedom, and in our case with $k=2$, the 90\%\,C.L. critical value is therefore $\chi_{\mathrm{crit}}^2=4.61$. In case of a boundary-constrained parameter space like $g_\nu^2 \geq 0$, Wilks' theorem may not apply exactly \cite{Cowan2011-pt}. For the specific case we investigate here, Monte Carlo studies confirm that the nominal critical value leads to conservative (weaker) constraints in the absence of a signal. In case of a strong detectable signal, proper coverage and applicability of Wilks' theorem are recovered. In the following, we adopt $\chi_{\mathrm{crit}}^2=4.61$ for a conservative sensitivity estimation. In Fig.~\ref{fig:sensitivity_with_other_bounds}, the corresponding purely statistical 90\%\,C.L. sensitivity of our TRISTAN benchmark  setup for tritium beta decay is shown, together with constraints from complementary searches discussed in Sec.~\ref{sec:Other}, at masses $m_a \geq 1\,\mathrm{keV}$. This corresponds to the parameter space relevant for our UV model in Sec.~\ref{sec:Model}. Below $m_a \sim 1\,\mathrm{keV}$, the sensitivity is nearly constant, as the shape analysis of the full spectrum becomes insensitive to the ALP mass. Our updated treatment -- both in the theoretical description of the spectrum and the more realistic evaluation of statistics and analysis range -- leads to a significant modification of the sensitivity bound compared to the estimate in \cite{Arcadi:2018xdd}.

\section{Complementary searches}\label{sec:Other}

The same interaction of \eqref{int}, and thus the emission of a neutrinophilic ALP, can also occur in other processes, ranging from laboratory experiments to cosmological and astrophysical environments. These processes are expected to provide additional constraints on the coupling $g_\nu$. In this section, we discuss the relevant constraints for a keV neutrinophilic ALP, revisiting some of the constraints from the previous study in \cite{Arcadi:2018xdd}.

\subsection{Meson decay}\label{sec:Meson}

The presence of a light neutrinophilic ALP induces new decay modes of SM particles, including the weak bosons, the tau lepton, and various mesons \cite{Lessa:2007up,Pasquini:2015fjv}. For the keV ALP considered here, with $m_a\ll m_K$, the strongest bound on the ALP--neutrino coupling $g_\nu$ comes from kaon decays measured by the NA62 experiment \cite{NA62:2020mcv,NA62:2021bji}. The bound was overestimated in \cite{Arcadi:2018xdd}, most likely due to a numerical error. Revising the bound, including one-loop contributions, the constraint on the coupling reads \cite{Dev:2024ygx} 
\begin{align}
g_\nu < 4.3 \times 10^{-3}~.
\label{eq:mesonBound}
\end{align}
This is the strongest laboratory bound on the general ALP--neutrino coupling, with a sensitivity comparable to that of tritium beta decay as illustrated in Fig.~\ref{fig:sensitivity_with_other_bounds}.

\subsection{Neutrinoless double beta decay}\label{sec:0nubb}

If the neutrinophilic ALP couples to a lepton-number-violating current, e.g. $\overline{\nu^c}\gamma_5\nu$, it induces additional channels for neutrinoless double beta decay ($0\nu\beta\beta$) \cite{Blum:2018ljv,Brune:2018sab}. The latest constraint from the EXO-200 dataset \cite{Kharusi:2021jez} reads
\begin{align}
g_{\nu,\slashed{L}} < 4-9 \times 10^{-6},
\end{align}
which is much stronger than the bounds from meson decays and tritium beta decay. Therefore, to allow the tritium beta decay experiment to provide the leading constraint, we focus on the scenario in which the neutrinophilic ALP couples only to the lepton-number-conserving current.

\subsection{New light degree of freedom and $N_{\rm eff}$}\label{sec:Neff}

The existence of new light degrees of freedom may also affect the cosmological evolution. If such particles are sufficiently light and interact strongly enough with the SM particles, they can reach thermal equilibrium with the SM plasma and contribute to the radiation energy density, thereby modifying the standard cosmological evolution.

For the mass range of interest, the keV ALP remains relativistic during the Big Bang Nucleosynthesis (BBN) epoch, giving rise to an effective contribution of $\Delta N_{\rm eff}\approx 0.57$, which is strongly constrained. Using PRyMordial \cite{Burns:2023sgx} to perform the BBN calculation, with the helium and deuterium abundances adopted from the PDG \cite{ParticleDataGroup:2024cfk}, we obtain a stringent bound, $\Delta N_{\rm eff}<0.30$ at 95\% C.L. from BBN. Therefore, we require that the new degrees of freedom not be thermalized at $T\approx1$ MeV so as not to affect the BBN epoch. This criterion allows us to derive a bound on the ALP--neutrino coupling $g_\nu$ \cite{Huang:2017egl}.

We first consider the scenario with a lepton-number-violating (LNV) current. The criterion described above places a constraint on the production cross section and thus on the coupling. For a keV neutrinophilic ALP, the process $\nu \bar{\nu} \to a$ is the dominant production mechanism. Since the production rate is proportional to the mass $m_a$, we obtain a constraint that scales inversely with $m_a$ as
\begin{align}\label{Neff_LNV}
g_{\nu,\slashed{L}} < ~2.2 \times 10^{-7}\left(\frac{1~{\rm keV}}{m_a}\right)~.
\end{align}
In addition, the process $\nu\bar{\nu} \to aa$ also contributes, especially for lighter ALPs, giving a mass-independent constraint of $g_{\nu,\slashed{L}} < ~\mathcal{O}(1) \times 10^{-6}$.

For the ALP coupling to the lepton-number-conserving (LNC) current, the production mechanism mentioned above is absent. Instead, the light degree of freedom most relevant for $N_{\rm eff}$ becomes the RH neutrino, which has the same mass as the LH neutrino in this scenario. It can be populated through the process $\nu_L \bar{\nu}_L \to \nu_R \bar{\nu}_R$, where the ALP acts as the force mediator. 
The criterion then yields
\begin{align}\label{Neff_LNC}
g_{\nu} < ~4.6 \times 10^{-6}~,
\end{align}
which is similar to the constraint from $\nu\bar{\nu} \to aa$, since both are $2\to2$ processes, and surpasses the still relevant bound from light ALP production in the early universe from the thermal plasma. Together with Eq. \eqref{Neff_LNV}, the two constraints are summarized in Fig.~\ref{fig:sensitivity_with_other_bounds} as $N_{\rm eff}$ (LNV) and $N_{\rm eff}$ (LNC), respectively.

For a lighter ALP with $m_a \ll 1$ keV, the effect can persist until the epoch of recombination and modify the formation of the Cosmic Microwave Background (CMB), leading to even stronger constraints. Since this mass range lies well below our region of interest, we refer the reader to Ref.~\cite{Sandner:2023ptm}, where the constraints are derived from the Planck data \cite{Planck:2018vyg}.

\subsection{Neutrino self-interaction}\label{sec:NSI}

Although the contribution to $\Delta N_{\rm eff}$ during the CMB epoch may be negligible, keV-scale neutrinophilic ALPs can still affect the CMB as well as Large-Scale Structure (LSS) through the additional neutrino self-interactions they mediate. Such interactions modify the evolution of cosmological perturbations by causing neutrinos to behave as a tightly coupled fluid rather than free-streaming particles, see \cite{Berryman:2022hds} for a review. Consequently, observations of the CMB and LSS constrain the strength of neutrino self-interactions. Interestingly, however, it has been shown that the current data can equally accommodate two distinct modes - “strongly-interacting” (SI$\nu$) neutrinos and “moderately-interacting” (MI$\nu$) neutrinos , corresponding to very different interaction strengths. In fact, some recent analyses even show a preference for strong neutrino self-interactions \cite{Poudou:2025qcx}.

Moreover, since the effects of neutrino self-interactions are degenerate with those of other $\Lambda$CDM parameters in cosmological fits, it has been pointed out that sizable neutrino self-interactions can help alleviate existing cosmological tensions, such as the Hubble tension \cite{Kreisch:2019yzn,Blinov:2019gcj}. Given that the underlying origin of these tensions remains unclear, we adopt the conservative bounds inferred from Ref.~\cite{Poudou:2025qcx}.

First, we consider the MI$\nu$ scenario, which places a stringent upper bound,
\begin{align}
g_\nu < 6.3\times 10^{-5} \left(\frac{m_a}{1~{\rm keV}}\right)~,
\end{align}
providing another strong constraint beyond those from laboratory experiments. However, if we instead consider the SI$\nu$ scenario, a coupling in the range
\begin{align}
g_\nu = 0.1-1.2\times 10^{-3} \left(\frac{m_a}{1~{\rm keV}}\right)~
\label{eq:SInu}
\end{align}
is still allowed, which is comparable to the current bounds from meson decays for keV-scale ALPs. Furthermore, if the ALP mass is below the CMB temperature, $m_a\ll T_{\rm CMB}$, the constraint becomes insensitive to the ALP mass and directly constrains the coupling. Using the latest CMB analysis from ACT \cite{ACT:2025tim}, the bound is
\begin{align}
g_\nu <  \mathcal{O}(1) \times 10^{-7}~.
\end{align}
However, this regime lies well below the mass range considered in this work.

\subsection{Supernovae}

The final constraint comes from astrophysics. Supernovae, as environments with extremely high neutrino densities, can probe neutrino interactions in a variety of ways. The emitted neutrinos can scatter off the cosmic neutrino background before being observed on Earth, thereby modifying the observed neutrino signal. Alternatively, the additional interactions can directly affect the supernova collapse, altering the expected supernova dynamics or even preventing the collapse altogether, see Ref.~\cite{Raffelt:2025wty} for a comprehensive review.

The precise predictions depend on the allowed processes ($\nu \bar{\nu}\rightarrow a, \, \nu \bar{\nu}\rightarrow a a, \, \nu \bar{\nu} \rightarrow \nu \bar{\nu}$), which in turn are related to the nature of neutrinos, as discussed in Sec.\,\ref{sec:Neff}. However, the parameter region of interest for KATRIN, corresponding to relatively strong neutrino interactions, remains unconstrained in both cases. For instance, \cite{Kolb:1987qy,Shalgar:2019rqe} find the requirement
\begin{equation}
    g_\nu < \mathcal{O}(1) \times  10^{-2}
\end{equation}
for keV neutrinophilic ALPs, while \cite{Heurtier:2016otg} constrain only couplings below $10^{-5}$ in the mass range of interest, since scalars with stronger couplings remain trapped in the core instead of escaping. In Fig.~\ref{fig:sensitivity_with_other_bounds} we show the latest constraint from \cite{Shalgar:2019rqe}. Furthermore, in \cite{Chang:2022aas}, it is predicted that large self-interactions can lead to the neutrinos forming a fluid flowing out of the supernovae, although \cite{Fiorillo:2023ytr} find that the effect on supernova dynamics is negligibly small.

\section{UV model for neutrinophilic axion-like particle}\label{sec:Model}

In this section, we construct a UV model that can generate the neutrinophilic ALP described in Eq.~\eqref{ALP} at low energies, while remaining relevant for laboratory searches. Moreover, we provide a detailed discussion of the thermal history that can evade the most problematic cosmological constraints discussed in the previous section. A benchmark model and an updated exclusion plot are also presented.

\subsection{Neutrinophilic ALP from an extended scalar sector}\label{sec:Scalar}

We start with the construction of a neutrinophilic ALP model from an extended scalar sector. Due to the strong constraints from neutrinoless double beta decay, we focus on a UV model based on Dirac neutrinos, with masses generated by the Yukawa coupling term $Y_\nu\overline L_L H^c \nu_R$. We consider a complex scalar $\phi$ which couples to the same Yukawa term through a dimension-five operator suppressed by a heavy new physics scale $\Lambda$, which is set to $\Lambda=30$ TeV in our benchmark, with further details given in Sec.~\ref{sec:FCC}. The neutrino mass term is then obtained from the UV Lagrangian as
\begin{align}\label{eq:UV}
-\mathcal{L}_{\rm UV}=\left(Y_\nu-\frac{\phi}{\Lambda}\right)\overline L_L H^c \nu_R+\text{h.c.}~
\xrightarrow{\text{$\langle H\rangle=v/\sqrt{2}$}}
\left( Y_\nu - \frac{\phi}{\Lambda} \right) \frac{v}{\sqrt{2}} \overline{\nu_L} \nu_R+\text{h.c.}~,
\end{align}
where the right--hand side shows the Lagrangian after electroweak symmetry breaking (EWSB), with the Higgs acquiring a vacuum expectation value (VEV) of $v=246$ GeV, which remains well above the scale of the new physics considered is this study.

Next, after the complex scalar $\phi$ acquires a VEV through spontaneous symmetry breaking with $\langle\phi\rangle =f_a/\sqrt 2$, we obtain the IR Lagrangian
\begin{align}\label{eq:IR}
-\mathcal{L}_{\rm IR}=\left(Y_\nu-\frac{(f_a+s)}{\sqrt{2}\Lambda}e^{ia/f_a}\right)\frac{v}{\sqrt{2}} \overline{\nu_L} \nu_R + {\rm h.c.}
\supset m_\nu
\overline{\nu_L} \nu_R
{-}\frac{v}{2\Lambda}(s+ia)\overline{\nu_L} \nu_R+ {\rm h.c.}~,
\end{align}
which contains two terms. The second term describes the interaction between the new scalar field, including the scalar ($s$) and pseudoscalar ($a$) modes, and the neutrino, which is the interaction we are interested in. We can then write
\begin{align}\label{eq:int}
\mathcal{L}_{\rm scalar}=g_\nu s\bar{\nu}\nu+ig_\nu a\bar{\nu}\gamma_5\nu\,,
\quad\text{where}\quad g_\nu=\frac{v}{2\Lambda}.
\end{align}
With $\Lambda\sim (10-100)$ TeV, we can obtain a coupling of $g_\nu\sim 10^{-2}-10^{-3}$, which is relevant for tritium beta decay experiments.

The first term gives the Dirac neutrino mass $m_\nu=M_\nu-\frac{f_av}{2\Lambda}$, where $M_\nu=Y_\nu v/\sqrt{2}$ is the bare Dirac mass term. The other mass term, on the other hand, is related to the new scalar, defined as $M_\nu^a=f_av/2\Lambda=g_\nu f_a$. The two contributions should nearly cancel, resulting in the observed tiny neutrino mass, which can originate from a non-trivial potential of $\phi$ preferring the minimum around $(\phi-Y_\nu\Lambda)\approx 0$.

In the IR, the scalar $s$ typically acquires a mass at the scale $f_a$. In our benchmark scenario, we set $f_a=100$ MeV and simply take $m_s=f_a$. By contrast, the pseudoscalar $a$, i.e. the neutrinophilic ALP, is a pNGB of the broken symmetry and can therefore be much lighter. We parametrize its mass as $m_a=\epsilon f_a$, where $\epsilon$ characterizes the amount of explicit symmetry breaking. For a keV-scale ALP, this corresponds to $\epsilon\sim10^{-5}$.

\subsection{Thermal history and supercooled phase transition}\label{sec:PT}

Next, we construct the details of the scalar $\phi$ phase transition that can fulfill the current constraints, especially the $N_{\rm eff}$ bound from the BBN epoch, which places a strong constraint on the existence of light degrees of freedom during $T \sim 1~\rm MeV$.

To satisfy the BBN constraint, we introduce an additional term involving the scalar and RH neutrino in addition to Eq.~\eqref{eq:UV},
\begin{align}
-\Delta\mathcal{L}_{\rm UV}=Y_R\left(f_a/\sqrt2-\phi\right)\overline{\nu_R^c}\,\nu_R~.
\end{align}
This term gives rise to a Majorana mass for the RH neutrino before the phase transition, making it sufficiently heavy, while the mass vanishes after the phase transition, leaving our Dirac neutrino scenario unaffected in the IR. With these ingredients in place, we then look into the details of the initial phase, the final phase, and the phase transition between them.

\paragraph{Initial Phase}

We start from the initial phase with $\langle\phi\rangle=0$. As we are considering the new physics at a scale around $100$ MeV, the heavy new physics can already be integrated out, and the electroweak phase transition is already complete, with the Higgs settled at $\langle H\rangle=v/\sqrt 2$. We can then focus on the extended scalar and neutrino sector.

For the scalar sector, we expect the mass of the degrees of freedom before symmetry breaking to be set by the scale $f_a$, so we simply take $m_s=m_a=100$ MeV for the real and imaginary components of $\phi$. 
For the neutrino sector, the mass terms after EWSB are given by 
\begin{align}
-\mathcal{L}_{M}=M_\nu\overline{\nu_L} \nu_R
+\frac{1}{2}M_R\,\overline{\nu_R^c}\,\nu_R+{\rm h.c.}~,
%=M_\nu\left(\overline{\nu_L} \nu_R+\overline{\nu_R^c} \nu_L^c\right)
%+M_R\,\overline{\nu_R^c}\,\nu_R
\end{align}
where we define $M_R\equiv Y_R\,f_a /\sqrt 2$. We expect the Yukawa coupling $Y_R$ to be of $\mathcal{O}(1)$ and thus $M_R\gg M_\nu \approx g_\nu f_a$, since $M_\nu-M_\nu^a=m_\nu \approx 0$. The LH neutrino therefore acquires a mass through the typical type-I seesaw mechanism, given by
\begin{align}
m_L=\frac{M_\nu^2}{M_R}=\frac{{ \sqrt 2}g_\nu^2}{Y_R}f_a\approx\frac{2.3}{Y_R}\,{\rm keV}\left(\frac{g_\nu}{4\times 10^{-3}}\right)^2\left(\frac{f_a}{100~\rm MeV}\right)~,
\end{align}
which is around the $1$ keV scale in our benchmark.

Under this setup, with the scale $f_a=100\,{\rm MeV}\gg 1$ MeV, the initial phase contains only new particles that are sufficiently heavy to either decay or annihilate before the BBN epoch, including the RH neutrino $\nu_R$ and the scalars $s$ and $a$. Only the LH neutrinos remain light, as in the SM, making this phase effectively equivalent to the standard cosmology. The LH neutrinos could become non-relativistic if the phase transition were not completed before the temperature dropped below $m_L\sim 1$ keV, as will be discussed later.

\paragraph{Final Phase}

After the phase transition is completed at the temperature $T_{\rm PT}$, the scalar $\phi$ acquires a VEV $\langle\phi\rangle=f_a / \sqrt 2$. The scalar mode $s$ is expected to remain heavy, with $m_s\sim f_a$, while the pseudoscalar mode $a$, being a pNGB, can be much lighter, with $m_a=\epsilon f_a$. Taking the explicit symmetry breaking factor $\epsilon\sim 10^{-5}$, as mentioned before, the ALP $a$ resides around the keV scale.

For the neutrino sector, once the scalar $\phi$ reaches $\langle \phi \rangle=f_a/ \sqrt 2$, the Majorana mass of the RH neutrino vanishes\footnote{No perfect cancellation is required, as long as the effective neutrino mass $m_{ee}$ probed by $0\nu\beta\beta$ decay is small enough \cite{deGouvea:2005er,deGouvea:2006gz}, the neutrino can also be pseudo-Dirac. Besides, although the mass term vanishes after the phase transition, the interaction between the scalar and the RH neutrino remains as $\frac{1}{\sqrt{2}} s e^{i a/f_a} \overline{\nu_R^c} \nu_R$. However, this interaction does not lead to relevant consequences, since the RH neutrinos are not populated.} and the neutrino becomes a Dirac fermion with mass $m_\nu = y_{\text{eff}} \, v / \sqrt{2}$ determined by the cancellation $y_{\text{eff}}\equiv Y_\nu - f_a/(\sqrt{2}\Lambda)$. We expect a fine cancellation between $Y_\nu$ and ${f_a}/{\sqrt{2}\Lambda}$ to satisfy the observed tiny neutrino mass, i.e. $y_{\text{eff}}\lesssim 10^{-12}$. Such fine-tuning could have its origin from a more evolved model, which is left for future study.

The comparison between initial phase and final phase is summarized in Table \ref{tab:BM1} with our benchmark values.

\begin{table}[t]
    \centering
    \begin{tabular}{|c|c|c|}
    \hline
         & Initial Phase ($T > T_{\rm PT}$)& Final Phase ($T < T_{\rm PT}$) \\
        \hline
        $\langle \phi \rangle$ & 0 & $f_a/ {\sqrt 2}$ \\
        $\nu$ nature & Majorana & (pseudo)Dirac \\
        $0 \nu \beta \beta$ & \cmark & \xmark \\
        $m_\phi$ & $m_s=m_a\sim f_a$ & $m_s \sim f_a , \, m_a \sim 1 \text{ keV}$ \\
        $m_\nu$ & $M_R \sim f_a$, $m_L \sim 1$ keV & $m_\nu\lesssim 0.45$ eV \\
        \hline
    \end{tabular}
    \caption{Benchmark Model ($g_\nu =4\times 10^{-3}$, $f_a = 100$ MeV, $\Lambda_\text{UV} = 30$ TeV) with properties before and after the Phase Transition from $\langle \phi \rangle=0$ to $\langle \phi \rangle=f_a/ {\sqrt 2}$.}
    \label{tab:BM1}
\end{table}

\paragraph{Supercooled phase transition and Thermal History}

If the phase transition process follows a generic thermal evolution, the phase transition temperature $T_{\rm PT}$ is expected to be around the $f_a=100$ MeV scale. A light ALP would then already be present before the BBN epoch and thus be subject to strong constraints.  Therefore, we need the phase transition to be supercooled, with the transition occurring only at a much lower temperature such that no light ALP is present before the end of the BBN epoch around $T\sim 30$ keV. 
Furthermore, there should not be a significant amount of reheating once the phase transition completes, as this would increase the temperature in the neutrino sector, thus modifying $\Delta N_\mathrm{eff}$. At the same time, to avoid the LH neutrino becoming non-relativistic 
which would change the cosmological evolution dramatically, we need the phase transition to be completed before $ T \sim  m_L\sim 1$ keV. Combining the requirements, we therefore consider $T_{\rm PT}\sim 10$ keV.

With the benchmark model mentioned previously, the cosmological evolution can be described as follows. In the early universe, at a high temperature $T>1$ GeV, the SM fields, the scalar $\phi$, and the RH neutrino $\nu_R$ are all in thermal equilibrium. When the temperature drops to $T\sim f_a=100$ MeV, everything except electrons, photons, and LH neutrinos has either decayed or annihilated away. At this point, there are no non-SM light degrees of freedom present, so the process proceeds as in standard cosmology. Then, around $T \sim 1$ MeV, first neutrinos decouple, obtaining a lower temperature than the photon bath, then electrons and positrons annihilate, and lastly big bang nucleosynthesis starts and is completed around $T \sim 30$ keV. The supercooled phase transition then takes place at $T\sim 10$ keV. After $\phi$ acquires a VEV, $a$ is a light pNGB, with its mass decreasing from $100$ MeV to $\mathcal{O}(1)$ keV, while remaining non-relativistic after the phase transition. Meanwhile, the neutrino becomes Dirac-like, and the RH neutrino emerges as a new light degree of freedom beyond the SM. Through the new neutrinophilic interaction, it reaches thermal equilibrium with the LH neutrinos. This process preserves the total energy density and therefore does not affect $N_{\text{eff}}$, which measures the overall invisible radiation. Instead, it could modify the neutrino temperature, as will be discussed in Sec.~\ref{sec:CNB}. The CMB spectrum, which forms only after the phase transition, is thus essentially identical to the SM prediction, with the only modification arising from neutrino self-interactions, whose strength is within current constraints.

\begin{figure}[tbp]
    \centering
    \includegraphics[width=0.9\textwidth]{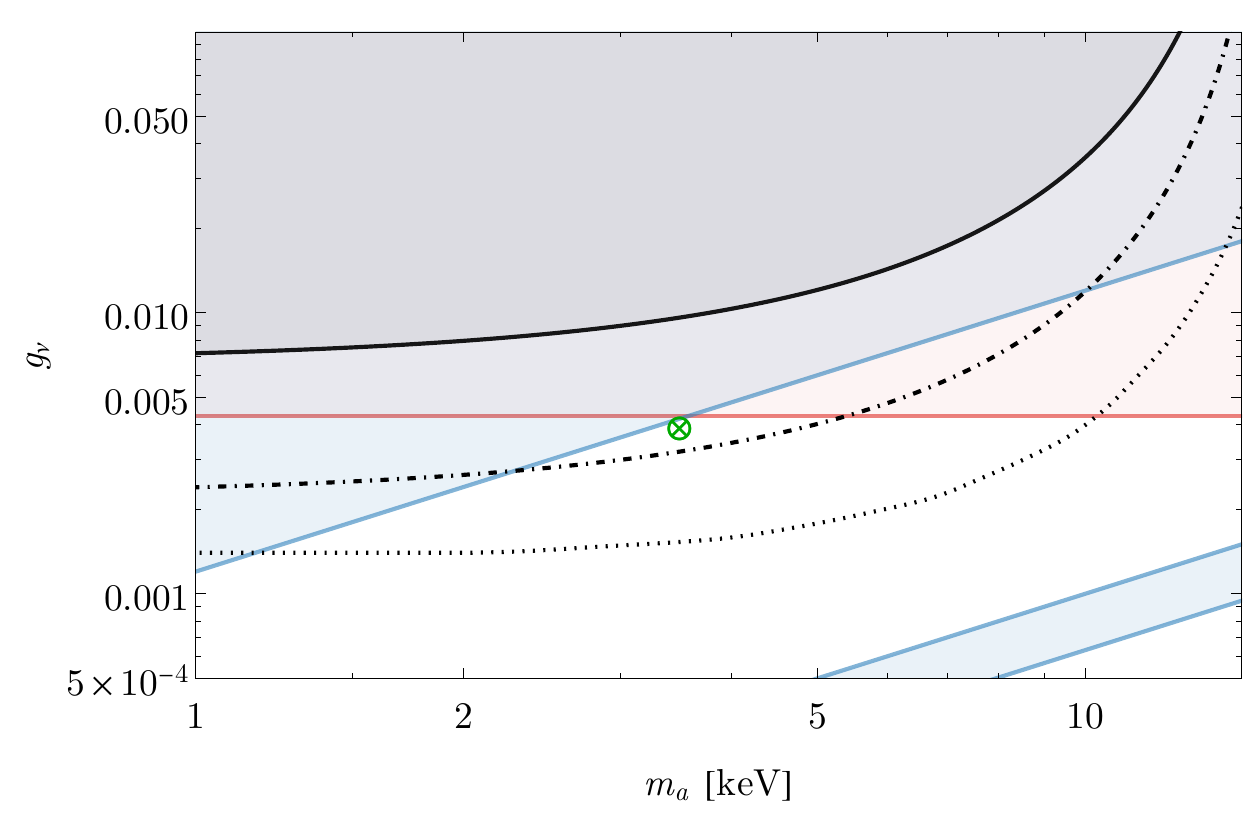}
    \caption{Relevant constraints and benchmark in the UV model. The blue shaded regions are excluded by neutrino self-interaction constraints, while the region in between represents the allowed parameter space considering the $\mathrm{SI}\nu$ mode (per Eq.\,\eqref{eq:SInu}). The red line shows the meson decay bound (Eq.\,\eqref{eq:mesonBound}) and the black line is the projected TRISTAN sensitivity benchmark described in Sec.\,\ref{sec:KATRIN}. As an outlook, we show $3\times$ the sensitivity as the dotted-dashed line, the minimal requirement to be sensitive to the model, as well as the dotted line which depicts the originally predicted sensitivity in \cite{Arcadi:2018xdd} ($\approx 7 \times$ updated sensitivity). The green cross is our benchmark point with a mass of $m_a = 3.5$ keV and a coupling of $g_\nu = 4 \times 10^{-3}$. }
    \label{fig:finalplot}
\end{figure}

\subsection{Revisited constraints in the UV model}

In Fig.\,\ref{fig:finalplot} we have reexamined the constraints on a keV-scale neutrinophilic ALP under the UV model we proposed, confronting them with sensitivity benchmarks for tritium beta decay. Now, the $N_\textrm{eff}$ bounds shown in Fig.\,\ref{fig:sensitivity_with_other_bounds} no longer apply, as during both BBN and CMB there is no additional radiation energy density. Instead, the most stringent bounds come from meson decay (in red) and neutrino self-interactions (for the strongly-interacting neutrino mode) from CMB measurements (in blue). The supernovae bounds have been omitted since they are not relevant in the regions of interest.

The benchmark, depicted by the green cross, is chosen as $m_a = 3.5 $ keV with a coupling of $g_\nu = 4 \times 10^{-3}$. We find that, to be sensitive to the relevant parameter space, at least a factor three of improvement in sensitivity (dotted-dashed line), corresponding to $3 \times 10^{16}$ electrons, is needed compared to the updated calculation with $4 \times 10^{14}$ electrons in Sec.\,\ref{sec:KATRIN} (solid line), where the statistical sensitivity scales as $g_\nu \propto N_\mathrm{tot}^{-1/4}$. The sensitivity originally estimated in \cite{Arcadi:2018xdd}, using $10^{18}$ electrons is shown as the dotted line. 

\section{New signatures from the UV model}\label{sec:other}

With a UV model at hand, additional particle states and new cosmological and astrophysical effects typically also arise. In this section, we briefly discuss potential new signatures that may be probed by future experiments.

\subsection{Searches for scalar mode}\label{sec:s}

Besides the light pseudoscalar mode $a$ that we focus on, the scalar mode $s$ also has a similar interaction to neutrinos as shown in Eq.~\eqref{eq:int}. Under our benchmark, the mass of the scalar after the phase transition is $m_s\sim \mathcal{O}(100)$ MeV. In this mass range, the cosmological and astrophysical bounds we mentioned above do not apply, and the main constraint comes from kaon decays, roughly resulting in the same bound as in Eq.~\eqref{eq:mesonBound}, but somewhat weaker because the mass $m_s$ is closer to the kinematic threshold. This also means that, if such a neutrinophilic ALP originates from a MeV-scale complex scalar $\phi$, once meson decay searches reach the corresponding sensitivity, we will not just see the signature from the emission of the light pseudoscalar, but also a mixed signal involving both the emission of the light pseudoscalar $a$ and the relatively heavy scalar $s$, with distinct energy distributions.

\subsection{Future collider searches}\label{sec:FCC}

Besides the light degrees of freedom, the UV model also features a heavy state responsible for generating the dimension-five operator ${\phi}\overline L_L H^c \nu_R/{\Lambda}$ introduced in Eq.~\eqref{eq:UV}. From the relation given in Eq.~\eqref{eq:int}, we find that the benchmark point $g_\nu=4 \times 10^{-3}$ corresponds to $\Lambda=30$ TeV, which implies the characteristic mass scale of this new heavy state.

There are three possible new states that can generate this dimension-five operator at tree level. They are simply the heavy versions of the particles participating in the interaction, including (1) a heavy Higgs doublet $H'$, (2) a heavy lepton doublet $L_L'$, and (3) a heavy right-handed neutrino $\nu_R'$.

Due to their heavy masses, the production and detection of these particles will likely require future colliders. With a center of mass energy of $\sqrt{s}=100$ TeV, the (1) heavy Higgs doublet $H'$ and (2) heavy lepton doublet $L_L'$ can be copiously produced through electroweak boson fusion. Their decays to the SM Higgs and leptons can also be readily detected. The (3) heavy right-handed neutrino $\nu_R'$ is more challenging to probe, as it is a SM singlet. Its production is possible through lepton-Higgs fusion but with a much smaller production cross section.

\subsection{Cosmic neutrino background}\label{sec:CNB}

For cosmological observables, the most important constraint usually comes from the measurement of $N_{\rm eff}$, especially when additional light degrees of freedom are present. In our case, however, the production of RH neutrinos occurs entirely within the already decoupled neutrino sector, so the total energy density of invisible radiation remains unchanged. Consequently, $N_{\rm eff}$, which measures the total energy density of invisible relativistic species, also remains unchanged.

However, the neutrino temperature does decrease due to this process. Although the total energy is conserved, the number of relativistic degrees of freedom increases, implying that the temperature of the neutrino sector, which now includes both LH and RH neutrinos, must decrease. Since the number of degrees of freedom doubles (from three to six), the resulting neutrino temperature becomes $T_\nu=T_{\nu,\rm SM}/\sqrt[4]{2}=1.64$ K, about $16\%$ lower than the SM prediction. When the cosmic neutrino background is eventually detected, such a deviation could provide evidence for the existence of RH neutrinos that interact efficiently with the LH neutrinos.

\subsection{Gravitational waves}\label{sec:GWs}

Since the model features a supercooled phase transition, it could be accompanied by a gravitational wave (GW) signal. Here, the phase transition is treated effectively, without determining its detailed dynamics -- a detailed analysis goes beyond the scope of this paper. Still, we can estimate the frequency at which GWs, if they are sourced, would be expected to lie. As shown in Sec.\,\ref{sec:PT}, the phase-transition temperature is expected to be $10$ keV, which corresponds to a Hubble scale $H \sim 10^{-19}$ eV. From this, the minimal GW frequency is given by $f = t^{-1} \sim H$ and then redshifted to today as $f_0 = a(t)/a(t_0) \,f$ \cite{Hindmarsh:2020hop}. Thus, here
\begin{equation}
    f_0 \gtrsim 10^{-13} \text{ Hz}
\end{equation}
would be expected, lower than for standard early-universe phase transitions and out of reach for e.g. LISA \cite{LISA:2024hlh} or NANOgrav \cite{NANOGrav:2020bcs,NANOGrav:2021flc}, requiring new methods to approach this parameter space.

\subsection{Not-quite-primordial black holes}
As proposed in \cite{Qin:2025ymc}, smaller density fluctuations than usually required to form primordial black holes (BHs) can still lead to the formation of dark matter halos, which then act as BH seeds and later directly collapse, a scenario they denote as ``not-quite-primordial BHs''. The direct collapse become possible because the CMB suppresses molecular cooling, leaving only atomic cooling via hydrogen. First order phase transitions, where the phase transition proceeds via bubble nucleation, inherently lead to density fluctuations. This can be understood as follows; the field tunnels stochastically at different times in different patches of the universe. The vacuum energy released during the phase transition and which is converted to matter or radiation therefore starts red-shifting at different moments, leading to a variation in energy density. Using the estimated $H_\mathrm{PT}\sim 10^{-19}$ eV for the $10$ keV phase transition in our model, the comoving scale is roughly given by $k \sim (a_\mathrm{PT} / a_0) \, H_\mathrm{PT} \sim 10^2 \, \mathrm{Mpc}^{-1}$, the same order of magnitude as used for a benchmark in \cite{Qin:2025ymc}. As in the previous section, a detailed analysis goes beyond the scope of this paper, and if the phase transition were to proceed sufficiently fast, the bubbles would not have time to grow to horizon size, therefore increasing the characteristic $k$, but it would be interesting to explore the possibility of forming not-quite-primordial BHs in future work. If such early BHs are formed, they can explain the little red dots observed by the James Webb telescope, act as seeds for supermassive BHs and explain the observation of early accreting BHs and early galaxy formation \cite{2023ApJ...954L...4K,2023Natur.616..266L,2023ApJ...959...39H,Maiolino:2023bpi,Matthee:2023utn,Maiolino:2023zdu,Bogdan:2023ilu,Natarajan:2023rxq,Perez-Gonzalez:2025bqr}.

\section{Conclusions}\label{sec:Conclusion}

In this paper, we have studied the phenomenology of neutrinophilic ALPs and their discovery potential in tritium beta decay experiments, focusing on the KATRIN experiment and its upcoming TRISTAN detector upgrade. Compared to the previous analysis in \cite{Arcadi:2018xdd}, we performed a more realistic evaluation of the experimental statistics and analysis window, together with an improved theoretical description of the beta decay spectrum. We also presented a detailed discussion of complementary searches and corrected estimates given in previous studies, providing the most up-to-date and reliable constraints on a keV neutrinophilic ALP for both lepton-number-conserving and lepton-number-violating interactions.

Based on our study of a benchmark model for KATRIN with the TRISTAN detector upgrade, we found that tritium beta decay experiments can achieve a sensitivity comparable to the current bounds from meson decays for the lowest ALP masses of a few keV considered here. A moderate increase in exposure beyond this benchmark statistics would allow tritium beta decay experiments to provide the leading laboratory constraints on lepton-number-conserving couplings that naturally arise in Dirac neutrino scenarios.
On the other hand, cosmological constraints remain much stronger, but they are known to be model dependent, relying on the details of the cosmological evolution. We therefore explored a UV model in which these constraints can be significantly relaxed. In particular, we considered the case where the ALP arises as a pNGB of a spontaneously broken symmetry featuring a non-trivial thermal history. If the symmetry is broken through a supercooled phase transition between the BBN and CMB epochs, the conventional cosmological bounds can be avoided and the searches in tritium beta decay experiments become relevant.

Furthermore, such a UV completion predicts additional experimental signatures. For the benchmark scenarios relevant to tritium beta decay, the required coupling and mass scale imply the existence of a scalar state with a mass around $\mathcal{O}(100)$ MeV, together with heavy new particles at the $\mathcal{O}(10)$ TeV scale. The required thermal history can also give rise to distinctive cosmological and astrophysical signatures, such as a lower neutrino background temperature.
Taken together, these correlated signatures provide a coherent experimental program for probing both the low-energy phenomenology and the underlying UV completion.

\acknowledgments

We thank Evgeny Akhmedov, Andrew Gavin, Sudhakantha Girmohanta, Ferenc Glück, Yann  Gouttenoire, Julian Heeck, Andrea Incrocci, Alexey Lokhov, and Edoardo Vitagliano for many useful suggestions and discussions. YC is supported by IBS under the project code, IBS-R018-D1. JL is supported by the Helmholtz Association and by the Ministry for Education and Research BMBF (grant numbers 05A23VK2 and 05A26VKB). The authors would like to express special thanks to the Mainz Institute for Theoretical Physics (MITP) of the Cluster of Excellence PRISMA+ (Project ID 390831469), for its hospitality and support during the completion of this work.

%\newpage

\bibliographystyle{jhepbst}
\bibliography{NuALP_ref}{}

\end{document}